\documentstyle[preprint,eqsecnum,aps]{revtex}
\tightenlines
 5
\begin{document}
\draft
\title{Nodal trajectories of spin observables
 and kaon photoproduction dynamics}
\author{Bijan Saghai}
\address{
Service de Physique Nucl\'{e}aire, CEA/DSM/DAPNIA,
Centre d'\'{E}tudes de Saclay, \\
F-91191 Gif--sur--Yvette, France
}
\author{ Frank Tabakin~\cite{byline} }
\address{
Department of Physics\& Astronomy,
University of Pittsburgh,
Pittsburgh, PA  15260
}

\date{\today}
\maketitle
\begin{abstract}
Spin observables for the reaction
 $\gamma p \rightarrow K^+ \Lambda$ are examined
 using three recent dynamical models and are compared to the
 general features of
such observables deduced earlier by Fasano, Tabakin and Saghai.
These general
features, such as the energy dependence of spin observables
and the location of nodes in their angle dependence,  are realized.
  Several instructive surprises,  which occur in this
comparison to the conjectures of Fasano et al.,  are then discussed.
The sensitivity of spin observables to isobar and {\it t-}channel
dynamics is analyzed and suggestions for selecting experiments which
provide important dynamical information are presented.

\end{abstract}
\narrowtext
\section{ INTRODUCTION}

The measurement of almost complete sets of spin observables has
become technically
feasible because newly developed polarized electron
or photon beams and polarized targets  offer high enough
luminosity to permit
measurement of the relevant observables.
The reaction $\gamma + p \rightarrow K^+ + \Lambda$
 is particularly advantageous,
since the angular asymmetry in the  parity nonconserving
weak decay $\Lambda \rightarrow p + \pi^-$
provides a direct measurement of the $\Lambda$'s polarization.
If the $\Lambda$'s spin
state is measured along with the polarization of the beam ($\gamma$)
or target (p),  then
spin transfer and spin rotation observables can be measured.
 Such spin-rich
experiments are underway at
 ELSA~\cite{ELSA} and planned at CEBAF~\cite{CEBAF} and
GRAAL~\cite{GRAAL}.

In an earlier publication~\cite{FTS}(FTS), the general structure of
 the full set of 16 observables for
K$^+$ photoproduction was examined.  In that discussion,   helicity
amplitudes proved to be particularly
useful for deducing general rules concerning the angular structure of
the 15 spin observables.
The sixteen observables (the cross-section plus 15 spin observables)
were found to fall into four
 ``Legendre classes,''  with four members
in each class.  The observables in each class have similar
``nodal structure'' possibilities  e.g.,
their values at 0$^{\circ}$ and 180$^{\circ}$ and their
 possible intervening nodes are of related nature.
 (That classification procedure, along with the advantages of
transversity
amplitudes, has led to a reformulation of the
general problem of determining which experiments
constitute a complete set of measurements.  A
  generalization to many reactions, including
 electroproduction and photoproduction of vector mesons,
 will be published separately~\cite{SPT}.)

In addition,  FTS used various truncations,
 both in a helicity basis and in a multipole representation,
 to deduce rules concerning the nodal structure
 and energy evolution of all 15
spin observables.  In this paper, those FTS
 rules are confronted with specific dynamical models.
The considered  models\cite{AS,WJC,SALY} are all based on an isobaric
approach using diagrammatic techniques. These models include extended
Born terms and a very limited number of resonances in {\it s-} and
{\it u-}channels supplemented by  contributions from the
{\it t-}channel;
namely,  by $K^*(892)(1^-)$ and $K1(1270)(1^+)$ as well
as $K^+$ meson exchanges.

We now comment briefly on each model and give the
relevant baryonic resonances.
First,  the model of Adelseck and Saghai(AS)~\cite{AS}
 is used to calculate the spin
observables.
 Their model,  which represents a good fit
  $(\chi^2/N_{d.f.}~\simeq~1.4),$   to the
 existing data up to $E_{\gamma}^{\rm lab}\leq1.5 $ GeV,
   uses the
$N^*(1440)1(1/2^+)$ and $\Lambda^*(1670)0(1/2^-)$ baryon
resonances.~\footnote{ The  quantum numbers
of the baryons
  are indicated as $\ell_{K \Lambda}(J^\pi)$.}
 In addition,  SU(3) constraints, based on their
success in strong interaction dynamics, are used to limit
 an otherwise oversupply of possible fits to the data. Then,
we also study the work
of Williams, Ji and Cotanch(WJC)~\cite{WJC},  who examined
electromagnetic production
processes with
photons(real/virtual) energies up to 2.1 GeV.
 That group included the N$^*(1650)0(1/2^-)$,
N$^*(1710)1(1/2^+)$, and $\Lambda^*(1405)0(1/2^-)$
baryon exchanges,  plus a significant incorporation
 of crossing symmetry requirements.

The third dynamical model considered here
 is a new model from the Saclay-Lyon group called SALY~\cite{SALY},
which also gives good
agreement with photoproduction data up to 2.1 GeV and satisfies
broken SU(3) symmetry requirements.
The SALY model includes all of the above AS baryon resonances plus the
 $N^*(1700)  2(3/2^-), $ $N^*(1720)  1(3/2^+),$  and
 $\Lambda^*(1600) 1(1/2^+)$ baryon resonances.
We emphasize that the first two models(AS \& WJC)  include only
spin-1/2 baryonic
resonances,  while the third one(SALY) also  includes spin-3/2
nucleonic resonances.
Spin 5/2 resonances are not included in any of these models.

  Among these three models,  the WJC model has the weakest
kaon exchange ({\it t-}channel) contributions, while AS has the
strongest. The WJC model produces weak {\it t-}channel exchange
because their main coupling constants
$(g_{K \Lambda N}$ and  $g_{K \Sigma N} )$
were not subject to SU(3)-symmetry constraints and in fitting data
they  generate quite small values of  both the
 {\it t-}channel and $g_{K \Lambda N}, g_{K \Sigma N}$
 couplings compared to those produced in the AS and SALY fits.
The {\it t-}channel coupling constants come out to be smaller in SALY
than in AS. This feature arises since the SALY model includes spin-3/2
baryonic
resonances and hence the need for the {\it t-}channel exchanges is
reduced,
in line with duality ideas~\cite{Collins}.
The role of  duality (the interplay between s- and {\it t-}channel
strengths)  and its effect on spin observables is
discussed in Section III.A.2.

Can the general rules for spin observables conjectured in FTS
and summarized in Section II,  be seen in the dynamical results?
What is the behavior of spin observables at lower
energies and what role is played by particular baryonic resonances
and by kaon exchanges?
How do the specific isobar and {\it t-}channel dynamics of the
three models affect spin observables?
Are some spin observables particularly sensitive to interesting
dynamics and are therefore  particularly important to measure?
  Those questions are  addressed  in Sections III \& IV.

\section{General Rules Revisited}

The general rules for the sixteen observables are described
 in detail in FTS~\cite{FTS}, which also
includes the derivation of such observables from a density matrix
approach.  Here we
simply highlight and confront those rules.  For convenience,  the
definition of the sixteen observables are recalled in
Appendix~A.

\subsection{The Helicity Rules}

The Legendre classes of the sixteen observables,
 which are labeled by ${\cal L}_0$, ${\cal L}_{1 a}$,
${\cal L}_{1 b}$, and ${\cal L}_2$, are:
$${\cal L}_0 ({\cal I}; \hat{E}; \hat{C}_{z'}; \hat{L}_{z'}),  $$
$${\cal L}_{1 a} (\hat{P}; \hat{H}; \hat{C}_{x'}; \hat{L}_{x'}), $$
$${\cal L}_{1 b} (\hat{T}; \hat{F}; \hat{O}_{x'}; \hat{T}_{z'}),$$
$${\cal L}_2 (\hat{\Sigma}; \hat{G}, \hat{O}_{z'}; \hat{T}_{x'}).$$
These observables, which are defined as the product of the usual spin
observable and the cross-section function ${\cal I},$ are
called  profile functions~\cite{FTS}.
The  profile functions are proportional to bilinear products
of amplitudes.~\footnote {The class ${\cal L}_2$ observables are
determined by the real and
imaginary parts of the bilinear combination of helicity amplitudes
$ ( \pm H^*_1 H_4 \pm H^*_2 H_3 );$ for
class ${\cal L}_{1a}$ observables by the real and imaginary parts of
$ ( \pm H^*_1 H_3 \pm H^*_2 H_4 ) ;$ and for
class ${\cal L}_{1b}$ observables by the real and imaginary parts of
$ ( \pm H^*_1 H_2 \pm H^*_3 H_4 ). $
Finally, the class ${\cal L}_{0}$ observables
are determined by the four combinations of the
magnitudes $ \pm |H_1|^2 \pm |H_2|^2 \pm |H_3|^2 \pm |H_4|^2 .$ }
In the above list, the first entry in each class is the cross-section
or a single
polarization observable
 $( {\cal I}, \hat{P}, \hat{T}, \hat{\Sigma});$
  the others are all double polarization
observables,  which appear ordered as Beam-Target
 $( \hat{E}, \hat{H}, \hat{F}, \hat{G}),$
Beam-Recoil
 $( \hat{C}_{z'}, \hat{C}_{x'}, \hat{O}_{x'}, \hat{O}_{z'});$
with the last entry in each class being the Target-Recoil
observables
 $( \hat{L}_{z'}, \hat{L}_{x'}, \hat{T}_{z'}, \hat{T}_{x'}).$
 The angular dependence of the
 above observables  are determined  by expressing the four
 helicity amplitudes $H_i (\theta)\  (i=1 \cdots 4)$ in terms of
 Wigner rotation functions, with $\theta$ denoting the produced kaon's
center-of-mass angle.
It is then simple to deduce that
${\cal L}_m$ class
observables are to be expanded in the associated Legendre,
 $P_{\ell m} (\cos \theta), $
functions. Thus,  ${\cal L}_{1a} and {\cal L}_{1b} $ involve sums
$\Sigma P_{\ell 1}, $ which
 vary as $ \sin \theta $.
 Hence, ${\cal L}_{1a} and {\cal L}_{1b}$  vanish at
 $0^{\circ}$ and $180^{\circ}.$
  Similarly,  class ${\cal L}_2$ spin observables involve sums
$ \Sigma P_{\ell 2}, $ which vary as $\sin^2 \theta$ and hence
 vanish  more sharply at those ``endpoints."  The ${\cal L}_0$ class
observables  are not necessarily zero at the ``endpoints."
However, all other spin observables
 must vanish at $0^\circ$
and $180^\circ.$

Following this helicity amplitude procedure, FTS deduced several rules:

\begin{itemize}
\item{\bf [h1] }{The observables $C_{z'}(\theta)$ and $L_{z'}(\theta),$
 as functions of the
scattering angle $\theta$, must have
 an odd number of sign changing~(SC) nodes,
 if they are nonvanishing observables.~\footnote{When a function passes
through zero we call it a node or more explicitly a sign-changing (SC)
zero.
If it touches zero without passing through zero,  we also call it a node
or a non-sign changing(NSC) zero.}  }
\item{\bf [h2] }{The observable $E(\theta)$ can be nodeless or have an
even number of sign-changing zeros.}
\item{\bf [h3]} {If the final-state  $J \geq 3/2 $ amplitudes vanish,
then the following relations must hold at all
angles:~\footnote{A sign misprint
 for $E$ and $H$ in Ref.~\cite{FTS} is corrected here. Also,
  the ${\cal L}_{1a}$ theorem in Section IV of FTS
is incorrectly stated as
having $T=H$ for no $J=3/2$ amplitudes;  the corrected
version is that $P=-H,$  when there are no $J \geq 3/2$ contributions. }
$$E \approx +1,\ \  L_{z'} = -C_{z'},\ \  L_{x'} = - C_{x'}\ \ P = -H.$$
  These conditions are expected to occur near threshold and to
rapidly change as the $J= 3/2 (K^+ \Lambda) $ P-waves turn on.
 Note that these observables are nonzero even when the
  $ J\geq 3/2$ amplitudes vanish.
These properties  follow directly from
 Eqs.~4.1-4.6 of FTS~\cite{FTS},  since
 only $H_1$ and $ H_3$ vanish if there are no
$J \geq 3/2$ amplitudes.~\footnote{The following relations
can be used to deduce these rules:
$\hat{L}_{z'}+ \hat{C}_{z'}=  |H_3|^2 - |H_1|^2;
\ \hat{L}_{x'}+ \hat{C}_{x'}= -2 {\rm Re}[ H_1 H_3^*];\ \
 \hat{P}+ \hat{H} =  -2 {\rm Im}[ H_1 H_3^*].  $ }  }
\item{\bf [h4]}{ The Legendre class ${\cal L}_{1a}$ profile
observables
 can have no more than $(2 J_{max} - 1)$ intervening nodes
 (i.e., not counting the end point nodes).  Here,  $J_{max}$
is the maximum $J-$value expected at a particular energy.
 Near threshold, these observables
are thus expected to be nodeless,
 with a node developing as the $J = 3/2$ amplitudes turn
on, but no more than 2 nodes being possible, until $J > 3/2$ states
appear.
 Note that the ${\cal L}_{1a}$ class observables are all nonzero even
if $J \geq 3/2$ amplitudes vanish. }
\item{\bf [h5] }{The ${\cal L}_{1b}$ class observables were expected
to be small near threshold, with nodes
developing only at higher energies,  since the $J \geq 3/2$ amplitudes
are expected to be small near threshold and
 these observables depend on interference
 between $J=1/2$ and $J\geq 3/2$ amplitudes.}
\item{\bf [h6] } { The ${\cal L}_2$
class observables should be small and  nodeless at lower energies.
 These observables also depend on interference
 between $J=1/2$ and $J \geq 3/2$ amplitudes.}
\end{itemize} We  shall refer to the above statements
as   FTS rules {\bf h1} $\cdots$ {\bf h6}.

These remarks are summarized in Table II of FTS.
  One interesting observation
is that ``if $\hat{C}_{z'}$ or $\hat{L}_{z'}$
 assume an even number of intervening zeroes,
then at least one of these zeroes would be of non-sign changing''
(NSC) type.
  This feature,  which will be
examined in our later comparison with dynamical models,
  is a consequence of the
requirement that both $\hat{C}_{z'}$ and $\hat{L}_{z'}$
 have an odd number of SC nodes -
which follows from basic parity requirements.  Similar remarks can be
made about $\hat{E},$ which must have an even number of SC nodes;
hence,  an odd number of zeroes means that there must be an odd number
of NSC zeroes.  Examples of these cases will be discussed later,
when bifurcations appear in the nodal trajectory plots.

  Since there are only four amplitudes
for pseudoscalar meson photoproduction,  only seven measurements
  of the
sixteen observables are needed to extract unique amplitudes
(one overall phase is arbitrary).  Some observables provide
redundant information and thus one needs to consult
the known rules~\cite{BDS} for selecting  seven independent
experiments.  That issue is not considered in this paper, see
Ref.~\cite{SPT}.

\subsection{The Multipole Rules}

Additional rules concerning spin observables  were
discussed by FTS, based on  the possible truncation of multipole
amplitudes.
The advantage of expanding the K$^+$ photoproduction amplitudes
 into multipoles
$E^{\pm}_{\ell}, M^{\pm}_{\ell}$ is that the
 orbital angular moment, $\ell,$ of the final
 $(K^+\Lambda)$ state can be used to  reduce the
number of
amplitudes,
 based on the existence of a centrifugal
barrier.
 Of course, this  truncation does not include the possibility of
 dynamical effects, which could magnify selected orbital
states. For example, a resonance could emphasize
a particular partial wave  or
competing effects could attenuate selected waves. However, it
is just the deviation from ordinary centrifugal-dominated behavior
 of spin observables that we
hope will serve as the best indicator of such dynamical effects.

Several additional features of the sixteen kaon
 photoproduction  observables
 were  conjectured in
FST, based on the suppression of higher orbital
angular momentum states.  These include:

\begin{itemize}
\item{\bf [m1] } A cross-section peak at 0$^{\circ}$ implies that
 the combination of the
P-waves defined by
$$\psi^{(1)}_P \equiv 3 E^+_1 + M^+_1 - M^-_1$$
(called type-1 splitting) and
the S-wave amplitude $E^+_0$ have
a relative phase angle of less than 90$^{\circ};$
provided that: (1) $\ell\geq$2 amplitudes are negligible,
 and (2) $\mid \psi^{(1)}_P \mid \neq 0$.  That is,
to get a cross-section peak one needs not only P-waves
  to interfere with the S-wave multipole,  but also the
   P-waves must have a nonzero (type-1) splitting.
\item{\bf [m2] }
A second type of P-wave splitting is defined by the
 linear combination of $\ell=1$ multipoles:
$$\psi^{(2)}_P \equiv 3 E^+_1 +  M^+_1 + 2 M^-_1 $$
(called type-2 splitting).  If $\ell \geq 2$
waves can be neglected and there are P-waves, but they
have zero type-2 splitting
  e.g.,  $\psi^{(2)}_P \rightarrow 0,$ then
(1) the spin observable  $\hat{T}_{z'} \rightarrow 0 $    at
$  90^\circ$
and (2) $\hat{P}$ and $\hat{\Sigma}$ are zero at all angles.

\item{\bf [m3] } The type-2 splitting also leads to the
following possible $\hat{P}$ behavior.
In order for the final $\Lambda$ polarization, $\hat{P}$, to have nodes,
 there must be nonzero type-2 P-wave splitting, $\psi^{(2)}_P \neq 0,$
  and  $\psi^{(2)}_P$ must not be
 collinear with $M^-_{1},$ unless  $ \ell \geq 2$ multipoles contribute
significantly.  Furthermore,  if the type-2 P-wave splitting is
nonzero and collinear with the $E^+_0$ multipole,  then, in the
absence of $\ell \geq 2$ amplitudes,  the
polarization $\hat{P}$ has a SC node at $90^\circ.$
\item{\bf [m4]}  Based on a scattering length expansion, it was
anticipated that $\hat{\Sigma} $  and  $ \hat{G}$ are small and nodeless
near threshold, with nodes developing only if
$\ell > 1$ amplitudes contribute.

\item{\bf [m5] } The beam-recoil observable $\hat{O}_{z'}$  and the
target-recoil observable $\hat{T}_{x'}$ are both zero
at all angles, unless
there are $\ell \geq 1$ multipoles and the stretched electric
and magnetic multipoles are unequal  e.g., $E^+_1 \neq M^+_1$ in
magnitude and phase.

\item{\bf [m6] } The observables $O_{z'}$  and $T_{x'}$ are
both of Legendre class ${\cal L}_2$ and are complementary
in their nodal behavior, see Eqs. 5.21 and 5.23 of FTS~\cite{FTS}.
By complementary we mean that if one tends to have a node, the other
does not so tend.

\item{\bf [m7] }
For $\hat{T}$ and  $\hat{F}$ to have SC nodes,
 type-2 P-wave splitting must be nonzero,  until
$\ell \geq 2$ waves contribute significantly.

\item{\bf [m8] } If the
Legendre class ${\cal L}_{1a}$ observables
 $C_{x'}$ and  $L_{x'}$ have zeroes then, as the momentum increases,
 these zeroes tend to be placed symmetrically about $90^\circ$.

\item{\bf [m9] } Near threshold, the class ${\cal L}_0$ spin observable
$C_{z'}$  has a SC node at $90^\circ.$

\item{\bf [m10]} Near threshold, the class ${\cal L}_0$ spin observable
$L_{z'}$  has a SC node at $90^\circ.$
\end{itemize}We  shall refer to the above statements
as FTS rules: {\bf m1} $ \cdots\ ${\bf m10}.
We are now ready to confront these FST rules with
the spin observables found using three different
dynamical models.

\section{The Dynamic Results}

The  kaon photoproduction observables are shown
for the Adelseck-Saghai(AS)~\cite{AS},
 Williams, Ji and Cotanch(WJC)~\cite{WJC},
 and SALY~\cite{SALY} models in
 Fig.~1  for $E_{\gamma}^{\rm lab} = 0.920 \ $GeV,
  and in Fig.~2  for $E_{\gamma}^{\rm lab} = 1.40 \ $GeV.
 The observables versus angle $ \theta$, {\it not} the profile
functions,  are presented in Figs.~1 \& 2. As emphasized in the
introduction, these models are based on
 selected baryon resonances and exchanged mesons,  with
  the physical basis for these models
  presented in Refs.~\cite{AS,WJC,SALY}.
  Here,  we simply compare these recent
dynamical models with the  FTS rules
{\bf h1} $\cdots$ {\bf h6 }  and    {\bf  m1} $\cdots$ {\bf m10}.

Note that these plots are organized  with  observables
of a given
Legendre classes$({\cal L}_0, {\cal L}_{1a}, {\cal L}_{1b}, {\cal L}_2)$
 in columns,  with
   the  first row  giving  the cross-section
and the
single spin observables $P, T $ and  $\Sigma.$
 The second row of plots in Figs.~1 and~2 gives the Beam-Target
$(E,H,F,G)$; the third row of plots   shows the Beam-Recoil
$(C_{z'}, C_{x'}, O_{x'}, O_{z'}).$
The fourth row of plots gives the  Target-Recoil
$( L_{z'}, L_{x'}, T_{z'}, T_{x'})$
  double spin observables.

  To facilitate comparison with the  FTS rules,
 the c.m. nodal angles (in degrees)
for the three models are plotted versus the
incident photon laboratory energy in Fig.~3.
We call these ``nodal trajectory" plots.
These curves have been calculated in the appropriate
energy domain for each model, i.e.
$E_{\gamma}^{\rm lab} \leq 1.5 $ GeV for AS and
$E_{\gamma}^{\rm lab} \leq 2.1 $ GeV for WJC and SALY.
The nodal angles are the angles  at which a
spin variable has a sign changing (SC) zero.  A single node which
moves with increasing beam energy appears
as a single curve.  A bifurcating curve shows the
energy evolution of a non-sign changing zero
into two SC nodes (see the observable $E$ for
a clear example).  These nodal plots are organized
by Legendre class and also as:
single spin observable(top row),  double-spin
Beam-Target(second row),  Beam-Recoil(third row) and
 Target-Recoil(bottom row).
  Examination of Figs.~1--3  allows one to confront the
three dynamical results with the FTS rules.

\subsection{Helicity rules}

  The observables of Legendre class
 ${\cal L}_{1a}, {\cal L}_{1b}$ and ${\cal L}_2$
are all seen to vanish as expected at the endpoints
 (0$^{\circ}$ and 180$^{\circ}$).
Only the set ${\cal L}_0 ({\cal I}; \hat{E}; \hat{C}_z; \hat{L}_{z'})$
 are nonzero at both of these endpoints.  Also note that the
${\cal L}_{1a,1b}$ class observables approach the end point angles
with nonzero slopes;  whereas, the
${\cal L}_2$ class observables approach the endpoint angles
with zero slope. Thus the ${\cal L}_{1a,1b} \propto  \sin\theta,$ and
 ${\cal L}_2 \propto \sin^2\theta$ properties are clearly seen in
Figs.~1 and 2.

\subsubsection{\bf Near and above threshold  results }

  Spin observables for the three dynamical models are now examined
 at two energies, $E_{\gamma}^{\rm lab} =$ 0.920 GeV and 1.4 GeV
for the purpose
of testing the FTS rules.
 The spin variables at $E_{\gamma}^{\rm lab} = 0.920$ GeV,  which are
near the $ (\gamma, K^+ )$ threshold of
$E_{\gamma}^{\rm lab} = 0.911 $ GeV, are shown in  Fig.~1. The case
of $E_{\gamma}^{\rm lab} = 1.4 $ GeV is shown in  Fig.~2.

 At 0.92 GeV,  both $C_{z'}$ and $L_{z'}$ display the anticipated sign
changing nodes,  see the FTS rules {\bf h1}, {\bf m9} and {\bf m10}.
The function $E$ is nodeless at low momenta and falls slightly
 below 1 for all three models,  which indicates
 that only small $J=3/2$ amplitudes
are in effect, see rules {\bf h2} and {\bf  h3}.
  Thus, the FTS suspicion that
 $E(0^\circ) \approx 1$ is an indication of small $J  = 3/2,$
 aligned $P$-waves is realized near threshold in these models.
 Note the enlarged  scale for the $E$ plot in Fig.~1.

The consequences of assuming small aligned $P$-wave amplitudes
 e.g., $ L_{z'} = - C_{z'},$  (rule {\bf h3}),
are realized for all three models at 0.92 GeV.  For example,  we also
see that  $ L_{x'} = - C_{x'}$ and  $P(\theta) = - H(\theta)\   $
 in Fig.~1.  Thus the rule {\bf h3} is fully realized near threshold.
 Indeed, the small deviation from these FTS rules can be used as a
measure of the  special sensitivity of each
spin observable near threshold to $J = 3/2,$ aligned P-wave amplitudes.

Assuming that at 0.920 GeV the ${\cal L}_{1a}$ observables are
dominated by $J=1/2,$  Fig.~1 shows  that the AS model has $J=1/2$
amplitudes of the largest magnitude,  that the WJC amplitudes are
smaller and  the SALY model gives in general the smallest $J=1/2$
contributions.
 The ${\cal L}_0$ observable $E$ also shows this 1/2 strength
pattern. This behavior is consistent with the coupling
constants~\cite{SALY} of the three models.  Note that the fact that
$P$ and $H$ are small compared to $C_{x'}$ and $L_{x'}$ indicates
constructive interference between the helicity amplitudes
$H_1$ and $H_4 $ e.g., they tend to be parallel near threshold.

The next FTS rule {\bf h4}, which is
 based on general helicity amplitude considerations, concerns the
maximum number of nodes for observables in Legendre class
 ${\cal L}_{1a} (\hat{P} ; \hat{H} ;
\hat{C}_{x'} ; \hat{L}_{x'}).$  For just $J=1/2$
amplitudes,  this class of observables should be nodeless
near threshold,  which is indeed the case at 0.920 GeV, see Fig.~1.
 As the $J = 3/2$ amplitudes turn on,  due to either
aligned P-waves (1 + 1/2 = 3/2) or unaligned D-waves
 $(2 -  1/2 = 3/2 ),$
 the ${\cal L}_{1a}$ observables could develop intervening nodes, but
not more than two,  until the $J > 3/2 $ states turn on.

  Indeed, at 1.4 GeV some ${\cal L}_{1a}$ spin observables do develop
nodes, see Fig.~2.
 In particular,  the
observables $P$ and  $H$,    each
 develop one SC node for the SALY model. For the SALY and WJC models
 $L_{x'}$ develops two nodes, while for AS, $L_{x'}$ stays at  one
node at 1.4 GeV. These results comply
with the FTS restriction {\bf h4}  that ${\cal L}_{1a}$ observables
have no more than two nodes until $J > 3/2$ amplitudes are strong.
 Hence, at 1.4 GeV the $J>3/2$ amplitudes are not explicitly seen in
the nodal structure of the ${\cal L}_{1a}$ observables.
The double nodal structure of $L_{x'}$ is a possible
 indication of interesting $J= 3/2$ dynamics in the
proton to $\Lambda$ spin rotation function ($D^{p \Lambda}_{L S},$
see Appendix A).

For the ${\cal L}_{1b} $ observables $(T; F; O_{x'}; T_{z'})$ at
0.92 GeV (Fig.~1), two are nodeless $(T,F)$
 and two $(O_{x'},T_{z'})$ have nodes in the vicinity of
$90^\circ.$
Thus the node part of rule {\bf h5} works for $T,F, $ but not for
$O_{x'}, T_{z'}.$
The ${\cal L}_{1b}$ observables depend on interference between
$J=1/2$ and $J=3/2$ amplitudes;  hence, we can conclude that these
$O_{x'}, T_{z'}$
nodes near $90^\circ$  yield important $1/2 \times 3/2$
 interference information for all three models. The reason for these
nodes is presented later.

For the ${\cal L}_{1b}$ observables at
 1.4 GeV (Fig.~2), $T$ acquires one node for WJC and SALY, while the
AS model remains nodeless.  The observable $F$  stays nodeless for
all three models at 1.4 GeV.  Double nodes appear in $O_{x'}$ for the
 SALY and WJC results, while AS stays at one node.
The above double nodes seem to be located symmetrically
 above and below $90^\circ.$
For $T_{z'}$   SALY gets three,
while AS and WJC stay with one node each, near $90^\circ$ and
$0^\circ,$ respectively.
  The mechanisms for these
changes with energy are more readily understood
by examination of the  nodal trajectory plots of the next section.

The ${\cal L}_2$  observables
$( \Sigma ;  G ,  O_{z'};  T_{x'})$ at 0.92 GeV ( Fig.~1)
comply with the rule {\bf h6};  they are all nodeless
and all, except perhaps $T_{x'},$ are small.
These observables,  as in the ${\cal L}_{1b}$ case, depend
on $J=1/2 \times 3/2$ interference,  which is why they are expected
to be small near threshold.
  At 1.4 GeV (Fig.~2), $\Sigma$
acquires an even number of nodes(2) for AS and WJC,  while  $\Sigma$
remains nodeless for SALY. Also $G$ and $O_{z'}$
each  acquire one node, but only for the SALY model.
The observable $T_{x'}$ remains nodeless at 1.4 GeV for all three
models,  but with much angular structure.
 All of these features agree with rule {\bf h6}.
However,  the acquisition of an even number of nodes in $\Sigma$ for
some models at and below 1.4 GeV is a surprise,  as will be discussed
later.

Now let us return to the
${\cal L}_0$ observables.
At 1.4 GeV (Fig.~2),  the ${\cal L}_{0} $  spin
observables  vary considerably with angle and some have nodes.   The
rules that: $E(\theta)$  must have an even number of nodes ({\bf h2}),
and $C_{z'}$  and $ L_{z'}$  an odd number of nodes ({\bf h1}),
   are clearly satisfied.
However, the three models manage to satify these theorems in
different ways, especially in the number of nodes.
This suggests that their ``nodal structure" and the
evolution of nodes with energy,
is a source of specific dynamical information.

 All of the models display nodal structures consistent
with the  {\bf h1} $\cdots$ {\bf h6} rules.
To understand how these nodes develop with increasing energy,
and the underlying reasons for the nodal structure of
spin observables,  it is more convenient to examine
 ``nodal  trajectory" plots.  We shall consider consequences
of  having a limited number of multipole amplitudes later.

\subsubsection{\bf Nodal trajectories}

To follow the detailed development of nodes with increasing energy,
we now examine  the ``nodal trajectory" plots of Fig.~3.
  The nodal angles (e.g., angles at which SC nodes occur)
are plotted versus the photon's laboratory energy.
  \\

{\bf (a) Single spin observables} \\

 Let us start with the
single spin observables ($P,T$ and $\Sigma$),  which are presented in
the top row of Fig.~3. (The cross-section,
and hence ${\cal I},$ is nodeless.)
   The recoil $\Lambda$
polarization, P in Fig.~3, has:  no nodes for the AS model;
 one node starting at $180^\circ$ at 1.47 GeV for the WJC case;
and an early node starting at $180^\circ$ at 0.96 GeV  for the SALY
model. Both AS and WJC models do not have spin 3/2 resonances,
in contrast to the spin 3/2 isobar that is part of the SALY model.
That isobar accounts for the dramatic difference in the observable
P;  namely, the early $180^\circ$ node for the SALY model and the
subsequent strong
energy dependence of its single node is a reflection of  the 3/2,
$\ell=1$ resonance   $N^*(1720)  1(3/2^+) $
at about $E_{\gamma}^{\rm lab} =1.1$ GeV, corresponding to the total
energy of $\sqrt{s} = 1.716$ GeV in the center-of mass frame.
 The other models rely on
{\it t-}channel exchanges for their $J=3/2$ strength at low energies.

 Similar reasoning
applies to
the observable $T$ in Fig.~3,  which also has an early $180^\circ$ node
for the SALY
model.  For WJC, the target polarization $T$ has a node
  due to  $J=1/2^+$ strength,  but the node displays
 smooth non-resonant evolution since the WJC model has no $J=3/2$
resonances. Thus, the nodal structure of $P$ \& $T$ are sensitive to
explicit {\it s-}channel spin 3/2 resonances,  but are not so revealing
concerning {\it t-}channel contributions.  The  curves $P$ and $T$ for
SALY are therefore good examples of resonance-driven nodal trajectories
and show how such plots can be used to extract detailed  resonance
dynamics.

The same SALY 3/2-resonances drive the spin observable $H$,  which  has
a  particularly dramatic nodal trajectory as seen in Fig.~3.  This case
will be discussed in the Beam-Target section.

  For the photon asymmetry observable, $\Sigma,$  the bifurcating
behavior seen in Fig.~3 for the AS and WJC models is unexpected.
This observable is {\it not required}
to have an even number of nodes.  (In contrast,   $E$ is
restricted to an even number of nodes,  which is the reason for its
 bifurcating nodal trajectory.)
{\it  To get two nodes in $\Sigma,$  especially at low
energies there has to be some J=5/2 amplitude strength.}  That strength
is apparently not
due to any 5/2 resonances,  since none of the models have
 explicit $J=5/2$ resonances.  Instead,  the $J=5/2$ strength at low
energies arises from a {\it t-}channel mechanism!

The polarized beam asymmetry $\Sigma$
 shows this unexpected double
nodal structure for the AS model at lower energies and at higher
energies for the WJC  model.  For the SALY model,  a single node
appears at $180^\circ$ and moves rapidly to $0^\circ;$   two nodes
appear at higher energy and evolve smoothly.
  In the region that SALY has one node,
this single node and its rapid motion can be accounted for
by its $J=3/2$ resonances.
The double node at higher energy for SALY occurs when it picks up
$J=5/2$ strength,  probably from   {\it t-}channel effects.
The general form of the $\hat{\Sigma}$ observable is
$\sin^2\theta ( a+ b~\cos\theta + c~\cos^2\theta),$
where
the terms $ a$ and $ b$ arise from interference between
$J\ge$ 1/2 states, while the term $c$ arises only if  $J\ge 5/2$
states contribute; for example,
$c$ arises first from a $3/2\times 5/2$ amplitude interference.
Therefore, to get the double nodal structure seen for the AS model near
threshold,  one needs a sizable $c$ term or, equivalently, at least
$J=5/2$ multipoles. It is also clear from this general form that for
small 5/2,  but sizable 3/2,  amplitudes,  the term $b$ arises first
from interference between
$3/2 \times 3/2$  and hence can give only  one
node.~\footnote{To generate $n$ SC nodes,   an observable
 needs to be described as
$\sin^m\theta $ times a polynomial of
 order $n$ in $\cos\theta.$  Here $m= 0,1,2$
depending on the Legendre class ${\cal L}_m.$ }

 The $J=5/2$ amplitudes could arise from a mechanism that boosts
the orbital angular momentum  to higher values.  The striking of
a virtual P-wave meson by a polarized incident photon (t-channel
exchange)  provides such a mechanism.
For example, if a virtual $K^+ {\rm or\ } K^*$ kaon
 peals off from the nucleon in a
P-wave,  it can  receive an extra orbital angular momentum boost when
it is struck by the incident photon.  The kaon then boosts
 to an $\ell =2$ state,  which, when added to
 the $\Lambda$'s  1/2 spin,  generates $5/2^-$ and $3/2^-$ strength,
even at low energies,  without an explicit 5/2 baryon resonance.
Thus, we learn that {\it t-}channel, or kaon exchange provides
$5/2^-$ and $3/2^-$ strength and therefore contributes to the
$(E^+_2, M^+_2)5/2^-$ and
$(E^-_2, M^-_2)3/2^-$  multipoles. This mechanism is illustrated in
Fig.~4.   The AS model  has particularly  strong
{\it t-}channel couplings,  which are generated,  in a duality sense,
to make up for
the absence of the 3/2 resonances contained in the SALY model.
The WJC model,  which has weaker coupling in
the {\it t-}channel processes,
 does ultimately also reveal a similar bifurcation in the $\Sigma$
observable,  but not until higher energies of  about 1.4 GeV.

Although the SALY model has stronger {\it t-}channel strength than
does the WJC model, it does not display the bifurcation one might
expect,  because it is dominated at low energy by the $J=3/2$
resonance. For the SALY model the above role for {\it t-}channel
$J=5/2$ amplitudes is not seen.  Instead, in Fig.~3,
the $\Sigma$ for SALY has a $180^\circ$ node which travels rapidly
toward $0^\circ$
in the region of the $J=3/2$ resonances of this model.  This property
is related to the resonance-driven
evolution of the single node in the observables $P,T$ and $H$ for the
SALY model.  In the $\Sigma$ case,
the 3/2 resonance increases the size of the $b \cos\theta$ term and the
rapid variation of $b$ with energy and its sign change
explain how the 3/2 resonance causes the
rapid $180^\circ \rightarrow 0^\circ$ nodal evolution.  At higher
energies the SALY model does
acquire 5/2 strength and thus a nonzero quadratic term,
$c\cos^2\theta,$
  permits the appearance of two nodes, as seen in Fig.~3.
Thus, all of the $\Sigma$ curves arise from resonance and/or
{\it t-}channel mechanisms.

We conclude that the observable $\Sigma$ provides a
particularly sensitive test of resonances and of $J=5/2$ amplitudes
and thereby of {\it t-}channel or meson exchange processes.  The basic
idea of duality, which equates a sum over all {\it t-}channel
processes with a sum over all {\it s-}channel resonances,  shows that
there is a close relationship between a dynamical model's content with
respect to assumed resonances and the corresponding strength of the
{\it t-}channel exchanges needed to fit the data.
Thus,  another way to describe the significance of
$\Sigma$ is that it tests the duality structure of the dynamics.
 This result and the enhanced role of $J=5/2$ amplitudes were not
anticipated by FTS.\\

{ \bf (b) Beam-Target spin observables} \\

We now discuss the nodal trajectories for the  Beam-Target observables,
which are given in the second row of Fig.~3.
For the observable $E,$  a bifurcation of nodes appears in Fig.~3.
Recall that $E$ is restricted to an even number of nodes by rule
{\bf h2}. Therefore,  if a (NSC)zero develops in this observable and
with increasing energy this vanishing extremum acquires a non zero
value, then it must  produce
two SC nodes: that is what we mean by a bifurcation.
 Insight into this rule is gained
 by noting~\cite{FTS} that $\hat{E}$
has the form $a +b \cos\theta + c \cos^2\theta.$    The coefficient $c$
is determined by
 amplitude products of the form  $3/2^+\times 3/2^+$ and interferences
$1/2^+ \times 3/2^+.$  Thus $c$
can exist either by having a $3/2$ resonance (as in SALY) or by enhanced
{\it t-}channel 3/2 strength,
without invoking 5/2 terms. Hence, the quadratic form can be realized
for those two reasons by all
models.  Indeed, all models (Fig.~3) display a bifurcation at about the
same photon energy;  for AS and WJC the term c is generated by the
kaon exchange process;  whereas, for the SALY case it is generated by
the underlying 3/2 resonances. There are no nodes in $E$ until about
 1.2 GeV,  even though the $a +b \cos\theta$ would seem to allow for
  low  energy nodes.  However, the $a$ term is dominated by
$|E^+_0|^2,$ while $b$ depends only linearly on $E^+_0$'s
  interference with $\ell=1$ multipoles.  Thus, the $a$ term dominates
and without a sizable contribution from $b \cos\theta,$
single nodes are prevented and one goes directly to double nodes
at higher energies.

The  Beam-Target observable $H$ is related to the single spin
observable $P$   which was discussed in the previous section.
The rapid variation of the nodal location  of $H$ for the SALY model is
due to the 3/2 resonance,  but
in a more dramatic fashion than for $P,$  see Fig.~3.
Thus the double spin observable $H$ has
enhanced 3/2-resonance dependence.
This is another good example of nodal structure generated by
resonances.  The relation between the nodal trajectories for $P$ and
$H$ is particularly instructive.
 We know that $ {\hat P} + {\hat H} = -2 $ {\rm Im}$(H_1 H^*_3) ,$
where the $H_1$ and $H_3$ helicity
amplitudes  are zero unless $J \geq 3/2$ amplitudes exist.
  Thus, in the absence of $J \geq 3/2$ amplitudes,
we recover the theorem $P = -H$ of rule {\bf h3}.
  In the case of the two  $J=3/2$ SALY-resonances,  the difference
between $P$ and $H$ is determined by    {\rm Im}$(H_1 H^*_3) $ and therefore
by the associated amplitudes interference.
The two SALY-resonances drive both $P$ and $H$ in a very revealing
manner.  Note that for the WJC model there
are no $J=3/2$ resonances and consequently nodes for that case develop
at higher energy and move from
$180^\circ$ to smaller angles smoothly for both $P$ and $H.$  This is
an example of a non-resonant driven
nodal evolution,  where the $J=3/2$ strength probably arises first
from {\it t-}channel effects.

Hence, observation of both $P$ and $H$ could reveal  the presence or
absence of resonance dynamics.
 One would need to measure the
polarization of the final $\Lambda$ {\it via} its decay
 and also measure $H$ which requires a linearly polarized photon
plus a proton target polarized in the $\hat{x}$ direction.
 In the present case,  we are dealing with two nearby
$J=3/2$ resonances albeit of different parity;
it would be of interest to explore the nature of
nodal trajectories for isolated and/or dominant resonances such as are
thought to occur in $\eta$ meson production~\cite{ETA}.

For $F$ we see no nodes until higher energies (above 1.4 GeV) for all
models.  The AS develops two nodes,
followed by WJC with two nodes and then SALY comes on with three nodes
in the 1.8 GeV region. This is clearly not driven by any of the
resonances,  they occur at lower energies. The form of $F$
is $\sin\theta( a + b \cos\theta + c \cos^2\theta + d \cos^3\theta),$
 where from Eq. D6 of FTS the $a$ term is enhanced by
the large $E^+_0$ multipole,  which makes it difficult for the
$b$ and $c$ terms,  needed to
generate  nodes,  to play a role.  But this amplification of the
$a$ term is also true for
the observables $P$ and $T$ and they do exhibit nodes;
  whereas, for $F$ there is a delay in the
appearance of nodes.  The $b$ term for $F$ must therefore
conspire to be smaller than for these earlier cases; it might be
characteristic of the observable $F$ not to have nodes only after
a critical energy is reached. When double
nodes do turn on
in $F $ for the AS model, it is because
$a$ and $c$ are comparable; we need to invoke $\cos^2\theta$
terms to get double nodes. To generate
$c,$ the AS model must use its considerable 5/2 amplitudes,
so again this suggests a {\it t-}channel mechanism.  The same reason
explains the later appearance of 2 nodes for the WJC model,
which is weakest in {\it t-}channel strength.  The SALY model
is the first to develop three nodes in $F$, which means the $d$ term
is effective,  but not until higher energies where $J \geq 5/2$
strength can be generated in several ways.

The observable $G$ has a single node for only the SALY model, as seen
in Fig.~3. This observable
is  of the form $\sin^2\theta( a + b \cos\theta ),$ where
$a$ depends on P-wave interferences of the type
($1/2^+ \times 3/2^+ + 3/2^+ \times 3/2^+),$ and also  S and D-wave
interferences.
  The $b$ term,  which is needed
to generate a single node, depends on P-and D-wave interference;
indeed, in $b$ the D-waves ($5/2^-$ and $3/2^-$ multipoles) appear
multiplied by P-wave amplitudes ($3/2^+$ and $1/2^+$ multipoles).
  Even above the resonance region,  P-wave amplitudes
 ($3/2^+$ and $1/2^+$) interfere with D-waves
   sufficiently to generate a $b$ that is comparable to $a$ only
for the SALY model,  which yields the single node in $G$ seen in Fig.3.
 For the AS and WJC model,
the $b$ term does not have sizable P-D wave interference;  also,
the S and D-wave interference keeps $a$ large enough,  due to large
S-waves,  to make it difficult in general to generate nodes in $G.$
A single node in $G$ provides evidence for significant
P-D wave interference and possible P-wave enhancement.

In many cases, especially at low energies,
   nodal trajectories are seen to reveal either
resonance or {\it t-}channel effects,
and/or enhancements due to  dominance of the $\ell=0$ multipole.
\\

{ \bf (c) Beam-Recoil spin observables} \\

Now consider the nodal trajectory plots for:
$\hat{C}_{z'}, \hat{C}_{x'}, \hat{O}_{x'}, \hat{O}_{z'}. $

At first glance, $\hat{C}_{z'}$ in   Fig.~3 seems impossible to
understand,  but it does have some
simple features. All models have a $90^\circ$ node near threshold.
This property can be understood from
 the general form for this observable   $\hat{C}_{z'}=
a + b \cos\theta + c \cos^2\theta + d \cos^3\theta,$ where
 near threshold $b$ is dominated
by   $|E^+_0|^2,$ while $a$ depends linearly on $E^+_0$'s
  interference with $\ell=1$ multipoles.
Thus near threshold the $\cos\theta$ term dominates and gives
the $90^\circ$ node near threshold.
The SALY 3/2 resonances show up in $C_{z'}$
 by rapidly moving that $90^\circ$ node
first to smaller and then to
larger angles.
   In the general form of the observable $\hat{C}_{z'}$
 the cubic term $d \cos^3\theta,$ appears even when a
truncation to $\ell \leq 1$ is used. The $d$ term involves interference
between the aligned  $J=3/2$ amplitudes.
Therefore,  in addition to explaining the rapid motion of the lower
energy node by 3/2 resonances,  at $\approx $1.31 GeV
 there is enough off-resonant
3/2 strength in the SALY amplitudes to invoke the cubic term and hence
yield the three nodes seen in Fig.~3.  Similarly,
 at $\approx$ 1.23 GeV  the AS model has
considerable 3/2 strength,
not by a resonance,  but by {\it t-}channel enhancement.  Thus, it
exhibits a smooth (non-resonant)
evolution of its low energy node, followed by an early turn-on of
three nodes.  This early turn-on
of 3/2 amplitude appeared earlier in the bifurcations seen in $\Sigma$
and $E.$  For WJC,  there
are only small 3/2 amplitudes (it has no 3/2 resonances and its kaon
exchange is small);  therefore,
it exhibits a smooth evolution of the $90^\circ$ node toward zero angle,

For $\hat{C}_{x'}$(Fig.~3) we note an absence of nodes until
  energies above 1.4 GeV. The AS
and then the SALY model exhibit one simple node,  while WJC turns
on with three above 1.61 GeV.
The prevention of early nodes is understood from Eq.~D9 of FTS, wherein
the leading $\sin\theta \times a$ term
is dominated by the S-wave multipole $|E^+_0|^2.$~\footnote{ Several
such cases occur where the multipole $|E^+_0|^2$  dominates the $a$
and  $ E^+_0 $ enters linearly in the $b \cos$ terms in
the polynomial.  That structure prevents nodes until higher energies.}

The spin observable $\hat{O}_{x'}$   has complicated nodal structure
that can be understood from its general form
 $\hat{O}_{x'}=
\sin\theta (a + b \cos\theta + c \cos^2\theta),$ where
 near threshold $b$ is dominant since it
 depends linearly on $E^+_0$'s interference with $3/2^+$ multipoles;
whereas, $a$ involves only $\ell \geq 1$ waves. Thus a $90^\circ$
node near threshold occurs for all three models, which although
not required by FTS rules, arises from dynamical dominance of S-waves
near threshold. That node moves most rapidly for the SALY model,
due to its 3/2 resonances;  the SALY model is also able to turn on the
$c$ term to generate two nodes above 1.26 GeV due to off-resonance
$3/2^+$ strength.  The AS model displays a smooth evolution of the
low energy $90^\circ$ node because it lacks a 3/2 resonance.  The WJC
exhibits structure in the
evolution of its $90^\circ$ node and also acquires two nodes above
1.2 GeV.  This WJC evolution arises perhaps from the energy dependence
of the S-wave amplitude and the onset of $3/2^+$ strength
and interference with $1/2^+$ amplitudes

For $\hat{O}_{z'},$  only SALY and WJC exhibit single and smoothly
evolving nodes. Its general form
 $\hat{O}_{z'}=
\sin^2\theta (a + b \cos\theta ),$ has
 $a$ dominant because it involves the S-wave $E^+_0$ multipole,  which
accounts for the delay in the onset of nodes,  which do occur when the
$3/2^+$ P-waves allow $b$ to compete with $a.$
\\

{ (d) \bf Target-Recoil spin observables} \\

Now consider the nodal trajectory plots in Fig.~3 for:
$\hat{L}_{z'}, \hat{L}_{x'}, \hat{T}_{z'}, \hat{T}_{x'}. $

For $\hat{L}_{z'},  $ recall that FTS {\bf h3} indicates that
$L_{z'} = - C_{z'}$ if $J\geq 3/2$ amplitudes vanish.
Similar nodal structure for these two observables is then expected
for models with small 3/2 amplitudes;  this holds true for
the AS case and, to a lesser extent, for the
WJC model.

 For $  \hat{L}_{x'},  $ the S-wave dominance of
 the $\sin\theta \times |E_0^+|^2$ term prevents low energy nodes.
By the time P-waves enter they are strong enough to invoke quadratic
terms and generate double nodes;  hence the bifurcation for all nodes
is seen. This is similar to the $E$ case, except there is no even node
 $\hat{L}_{x'}$ theorem.

The form of $  \hat{T}_{z'}  $
is $\sin^2\theta ( a + b \cos\theta + c \cos^2\theta);$
we have another case where the S-wave dominance of the $b$ term
produces $90^\circ$ nodes for all models near threshold,  followed by
a rapid variation for the SALY case due to its 3/2 resonance.  Once
the 3/2 amplitudes turn on at higher energies,  in part from
{\it t}-channel effects,  then the $c$ term enters and double
nodes appear.  This is similar to the $C_{z'}$ case.

Finally, with the form
 $  \hat{T}_{x'} =\sin^2\theta ( a + b \cos\theta )$ the S-wave makes
$a$ dominant and nodes are therefore postponed
until  high energies (1.5 GeV) for WJC and AS (just barely seen,
since this is the end of its region) and at 1.55 GeV for the SALY model.

Many of the characteristics described above are determined by the
dominance of S-waves and the subsequent appearance of P-waves.  The most
interesting cases are those that arise from isobar resonance  and/or
{\it t}-channel effects.

\subsection{Multipole rules--revisited}

 Each multipole includes reference to the final
 K$^+ \Lambda$ orbital angular momentum.
  Hence, explicit truncations are suggested and additional rules
{\bf m1} ${ \cdots }$ {\bf m10 }
 can be deduced for observables near threshold.  The assumption involved
in deducing these additional rules is that centrifugal barrier
suppression
of amplitudes dominates the dynamics, and that  resonance  or other
special dynamic effects can be neglected. However,  dynamical models do
include  baryon resonances  and particular {\it t-}channel exchanges and
therefore  it is expected that  the FTS  {\bf m1} $ \cdots $ {\bf  m10}
rules will be broken.  In contrast, the helicity rules   based
on general symmetry principles  are not and should not be broken.

To facilitate comparison of the dynamical results of AS, WJC, and SALY
to the rules offered by FTS, it is best to examine the Argand plots
 of multipoles(Appendix~B) for the AS, WJC and SALY
 models.  The $S$ and $P-$wave multipoles are displayed in Fig.~5a,
while the $D-$waves are in Fig.~5b.  Note the scale is fixed in Fig.~5a,
but varies with model in Fig.~5b with the AS having the smallest scale.
For convenience, the multipoles are phase rotated to give a real
$E_0^+$ multipole at theshold.  Only that $S-$wave multipole is nonzero
at threshold;  all others start at zero and evolve as
$q^\ell,$ where $q$ is the $K^+\Lambda$ linear momentum.  The WJC model
displays a rapid variation in $E^+_0$ due to its $N^*(1650)1/2^-$
isobar;
the two other models have $E^+_0$ multipoles that decrease without
resonance (counter-clockwise looping) structure.
Structure due to the $N^*(1710)1/2^+$ of WJC is seen
in the energy evolution of its $M_1^-$ multipole,  with all of its other
multipoles evolving nonresonantly.  The AS displays $M_1^-$ structure
from its $N^*(1440)1/2^+$ isobar.  The $M_1^+$ multipole has
 nonresonant structure in the AS
model.  For the SALY case,  that multipole displays resonance looping
due to the added $N^*(1720)3/2^+$ isobar,  which
also introduces some $E^+_1$ structure in the SALY model.  The SALY model's
$M_1^-$ also acquires resonant structure apparently due
to its $\Lambda^*(1600)$.

 The $D-$wave multipoles (Fig.~5b) are quite
different for the three models.  Only the SALY model
has explicit $D-$wave isobars and consequently that model displays
resonant $M_2^-$ and $E_2^-$ $3/2^-$ looping.  The $5/2^-$ multipoles
$(E_2^+, M_2^+)$ evolve smoothly from zero. The small AS
$D-$wave multipoles have some interesting structure,  which does not
arise from explicit $D-$wave isobars.  For the WJC model,  there
are no $D-$wave resonances and the multipoles evolve without structure.
Note that the ${\it t-}$channel mechanism illustrated in Fig.~4
feeds mainly into the $3/2^-$ and $5/2^-$ multipoles.
Thus,  aside from the SALY case, these $D-$wave multipoles
are connected to that mechanism.

The main properties of the Argand plots of Fig.~5 are the relative
strengths of the $\ell \leq 2$ multipoles and their relation to their
input resonances and the strength of their
{\it t-}channel mechanisms. The $\ell=0$ multipole $E^+_0$ is naturally
dominant near threshold.
It decreases smoothly for the SALY and AS models,  but due to the
isobars in the WJC model,  their $E^+_0$
multipole has a rapid looping at low energies.
Several characteristics are worth noting.

\noindent In the AS model:
\begin{itemize}
 \item{ the $(E^+_0)1/2^-$ multipole is driven
by the nucleon {\it s-}channel
term and a $\Lambda(1670)$
resonance;}
 \item{ the $(E^+_1, M^+_1)3/2^+$ multipoles are not driven by a
resonance nor by
{\it t-}channel exchange;}
 \item{ the $( M^-_1)1/2^+$ multipole is driven by one $N(1440)$
resonance;}
 \item{ the $(E^-_2, M^-_2)3/2^-$ multipoles are driven by
{\it t-}channel exchange;}
 \item{The $5/2^-$ multipoles $(E^+_2, M^+_2) $ receive
{\it t-}channel contributions.}
\end{itemize}
\noindent In the WJC model:
\begin{itemize}
 \item{ the $(E^+_0)1/2^-$ multipole is driven
by the nucleon {\it s-}channel
term and the
$\Lambda(1405)$ plus
$N(1650)$ resonances;}
 \item{ the $(E^+_1, M^+_1)3/2^+$ multipoles are not driven by a
resonance nor by
{\it t-}channel exchange;}
 \item{ the $( M^-_1)1/2^+$ multipole is driven by one $N(1710)$
resonance;}
 \item{ the $(E^-_2, M^-_2)3/2^-$ multipoles are driven  by
{\it t-}channel exchange,}
 \item{The $5/2^-$ multipoles $(E^+_2, M^+_2) $ receive
{\it t-}channel contributions.}
\end{itemize}
\noindent In the SALY model:
\begin{itemize}
 \item{ the $(E^+_0)1/2^-$ multipole is driven by the nucleon
{\it s-}channel term and by the $\Lambda(1670)$ resonance;}
 \item{ the $(E^+_1, M^+_1)3/2^+$ multipoles are driven by a
$N(1720)$ resonance;}
 \item{ the $( M^-_1)1/2^+$ multipole is driven by the two
$N(1440),\Lambda(1600)$
resonances;}
 \item{ the $(E^-_2, M^+_2)3/2^-$ multipoles are driven by a
$N(1700)$ resonance and by
{\it t-}channel exchange,}
 \item{The $5/2^-$ multipoles $(E^+_2, M^+_2) $ receive
{\it t-}channel contributions.}
\end{itemize}

 Using these Argand diagrams,  we can now discuss the FTS multipole
rules.
 The cross-section peaking rule of FST, under the
neglect of $\ell=2,$ requires that $\psi^{(1)}_P $ and $E^{+}_{0}$
be within $90^\circ $
when viewed as vectors in the complex plane.  When one plots
 $\psi^{(1)}_P$ and compares it to $E^+_0, $
 it is seen that the rule {\bf m1} holds true.
 Although many FTS multipole speculations do not occur because of the
{\it t-}channel mechanism,
 the rules {\bf m9} and {\bf m10} are true at low energies,  since
$E^+_0$
dominates and  $C_{z'}$ and $L_{x'}$ do have SC nodes at $90^\circ.$
 Other rules that hold true are {\bf m6}, which does get realized
in the complementary nature of the nodes; namely,  $O_{z'}$ has,
while  $T_{x'}$ does not have nodes at least below 1.5 GeV.
The other FTS multipole rules are not realized, because the FTS
truncation assumptions did not take into account that {\it t-}channel
exchange can  introduce higher $J$ amplitudes.

Using Fig.~5,  we can classify the
relative roles of the
multipoles for each model.
  For the three models, the relative importance of the
  multipoles, based on the maximum size of the
multipole over the full energy range,
 is given in Table I.  The resonant multipoles and those that
receive strength from {\it t-}channel exchange are also indicated.
The enhanced role for $D-$waves is generated by one or another
of these mechanisms depending on the dynamics of the model.

These multipole strengths are consistent, of course, with the driving
resonances and the
{\it t-}channel strengths of each model.

 The big surprise for FTS is
the importance of $\ell=2$
multipoles in the $3/2^-$ and $5/2^-$ states,  due to kaonic exchanges.

\section{Summary and Conclusions}

Based on general symmetry requirements, FTS~\cite{FTS} deduced general
rules for the 15 spin observables in the photoproduction of
pseudoscalar mesons. These rules, supplemented by assumptions
of smooth energy evolution and centrifugal barrier dominance
can be used to define the ``normal'' behavior of spin observables.
Deviation from some of these rules indicate a serious violation of
a symmetry, such as parity violation.  Deviations due to non-smooth
energy evolution,  or dominance of selected states,  are of dynamical
origin,  as in the case of underlying hadronic resonances.
As a prelude to analysis of future experimental results,
we have confronted the FTS analysis with  three
current models~\cite{AS,WJC,SALY}
  of the photoproduction of $K^+$ mesons.~\footnote{We have not
included the oft-quoted work of Ref.~\cite{Bennhold},
 since we found a serious error
in their code; namely, their $Q_{\ell}$ functions are wrong.  As a
consequence, their spin observables
display a large and incorrect number of nodes near
threshold~\cite{saghai1}.
  We appreciate receiving a copy of their code from C. Bennhold and
look forward
 to seeing their corrected parameters.  This was a case
where the general FTS threshold rules~\cite{FTS} served to detect an
error and shows one way  these rules can be useful.}

  All of the
predictions based on parity and angular momentum conservation
are realized in these models,  including statements about the
even or odd number of sign-changing nodes.  The only deviations
noted are those possibly attributed to special dynamics, such as
underlying resonances.  Indeed,  an important conclusion is that
observation of the nodes of spin observables,
as they unfold with energy, offers  a powerful way to
extract specific dynamical resonance information.

This conclusion is realized by addressing two crucial questions :
$\it 1)$ what can the nodal structure of the
{\it forthcoming polarization data} reveal
about the highest spin of the intermediate state baryonic resonances
required by the reaction mechanism? $\it 2)$ in {\it dynamical models},
how can we disentangle
the contributions due to genuine baryonic resonances from those
mimicked by the kaonic exchanges in line  with the duality hypothesis?

To summarize our findings, and in view of the envisioned
polarization measurements~\cite{ELSA,CEBAF,GRAAL} we single out
 our most significant results on the reaction mechanism deduced
 by confronting the FTS rules with specific models.

The $\Lambda$-polarization asymmetry is technically the easiest
to measured. Here, the nodal structure of $P$ (and also $T$) is
mainly of resonance-driven nature and hence is
sensitive to  {\it explicit s-}channel spin 3/2 resonances.
 The beam-asymmetry $\Sigma$ proves to be an appropriate
observable in testing the validity of the {\it duality hypothesis} in
the strangeness sector.   This duality
hypothesis is verified by investigating the underlying dynamics of
three models; wherein, 5/2 amplitude strength influences spin
observables $\it via$
  $3/2\times 5/2$ and/or $5/2\times 5/2$ interference, even
though
explicit 5/2 resonances are not included in the models.  The 5/2
strength arises from {\it t-} rather than from {\it s-}channel
dynamics.
 The beam-asymmetry observable
is also very suitable in investigating the role of {\it explicit}
spin 5/2 resonances in the reaction mechanism.

The Beam-Recoil asymmetries are found very attractive mainly with
respect to their sensitivity to $J=3/2$ resonances. The richest
information is
embedded in the $C_{z'}$ observable, obtained using a circularly
polarized beam. The odd number of its nodes
are produced through $\it different$ and $\it distinguishable$
mechanisms according to the
dynamical ingredients of the models, e.g., the explicit presence
of spin 3/2 resonances {\it versus} the manifestation of the duality
through the {\it t-}channel exchanges. This property is also true for
the Beam-Target observable $E$. In addition, the cubic term in
$\cos \theta$ arises through the 3/2 amplitudes interferences.
The other Beam-Recoil observable
($C_{x'}$) with circularly polarized beam shows high sensitivity
to the $J=3/2$ aligned $P$-waves producing an {\it explicit} signal
for the $J=3/2$ resonances. The asymmetries $O_{x'}$ and $O_{z'}$
corresponding to a linearly polarized photon beam contain
important information on   $1/2\times 3/2$ interferences.
Even at low energies, the $O_{x'}$ observable shows deviations from
FTS rules for the three models because of {\it dynamical} effects.

The next double polarization family, Beam-Target, contains
some common features with the above asymmetries. Namely, the
observables connected to the circularly polarized beam, $E$ and
$F$,  have behaviour comparable  to  $C_{z'}$ and $\Sigma$,
respectively. For the linearly polarized beam observables the
situation is slightly different. The $H$ asymmetry manifests
characteristics similar to those of $P$, with {\it s-}channel
spin 3/2 resonance effects amplified because of an enhancement due to a
large $E^+_0$ multipole in the $\cos \theta$ term,
while the non-resonant driven nodal structure due to
duality induces a very different evolution. The $G$ asymmetry
is driven by an spin rich interference
between 1/2, 3/2, and 5/2 terms. Hence, the appearence of a
single node in this observable provides evidence for significant
$P\times D$ wave interference and possible $P-$waves enhancement.

The last set of observables, Target-Recoil
asymmetries, are characterized by the dominance of S-waves and
the subsequent appearance of P-waves arising from isobar
resonance and/or {\it t}-channel effects, similar
to the cases seen already for observables within
other families. This redunduncy in information content is of course
expected~\cite{BDS} from analysis of the number of independent
experiments.

Nodal angle
versus $E_{\gamma}$ trajectories,  based on {\it direct experimental
information}, rather than specific dynamical models,  should
 provide a powerful tool in pinning down the reaction
mechanism of the strangeness electromagnetic production processes
and, hopefully, in the search for missing resonances~\cite{misre}.

\acknowledgments
We wish to thank Drs.~C.~Bennhold, J.~C.~David, S.~A.~Dytman,
 R.~A.~Eisenstein, C.~Fayard, G.~H.~Lamot, and F.~Piron,
 for their help and encouragement.
We both wish to express our appreciation for the
warm hospitality during visits to
the University of Pittsburgh(B.S.) and to Saclay(F.T.).

\begin{appendix}
\section{Spin Observables Recalled}

The definition of the various spin observables are
provided in the literature~\cite{FTS}.  For convenience,
we present a brief discussion of the sixteen observables.

The differential cross-section is defined by:
$$ \sigma(\theta) = \frac{q}{k}  {\cal I}(\theta) ,$$
with $q$ the final and $k$ the
initial c.m. momenta.
Here we extract the angle dependent function ${\cal I}(\theta),$
which is used in FTS to define ``profile functions.''
These profile functions are denoted by
$\hat{X} = {\cal I} \cdot  X,$   for any spin observable $X.$
  The profile functions are  determined by bilinear products of
amplitudes and therefore are useful for
extracting amplitude information.

Of the sixteen observables,  one is: (1)  the cross-section
function ${\cal I}$; three are {\it single spin observables}:
(2) the
polarization of the produced $\Lambda, $  $\hat{P};$  (3)  the
polarized target asymmetry,  $\hat{T};$ and (4) the photon
polarization asymmetry $\hat{\Sigma} .$

The remaining twelve spin observables are {\it double spin
observables}.
These are further classified as involving: (BT) polarized beam,
polarized target;   (BR) polarized beam, polarized recoil
$\Lambda$;  (TR) polarized target, polarized recoil $\Lambda$.
Each of these three types of double spin observables
have four members.  The (BR) type is also called a
spin transfer observable.
The (TR) type is also called a
spin depolarization observable, often denoted by the symbol D;
 for example,  the depolarization spin transfer variable
where the incident p and the final $\Lambda$ spin directions
are normal(N) to the scattering plane is usually called $D_{NN}$.
In the notation of FTS that  spin  variable is called
$C^{p, \Lambda}_{y, y'}.$ Their notation is based on the
spin correlation description,  where the superscript indicates
the two particles involved (e.g., the BT , BR, or TR classification)
and the subscript indicates direction.  These directions
are denoted by either the initial unit vectors,
$\hat{x},\hat{y},\hat{z},$
 or the final unit vectors, $\hat{x}',\hat{y}',\hat{z}'.$ Another
convention used is that of normal N,  sideways S,  and longitudinal L
directions:  $\hat{x} \equiv S,\hat{y} \equiv N,\hat{z} \equiv L.$

The relations between these spin-correlation (double spin)
observables
and the conventional set used here is for BT:
$$ \hat{E}  =  C^{\gamma, p}_{z, z} \ \ \ \
  \hat{H}  =  C^{\gamma, p}_{y, x}  $$
$$ \hat{F}  =  C^{\gamma, p}_{z, x} \ \ \ \
  \hat{G}  =  C^{\gamma, p}_{y, z} ;  $$
for BR:
$$  \hat{C}_{z'}  =  C^{\gamma, \Lambda}_{z, z'}  \ \ \ \
    \hat{C}_{x'}  =  C^{\gamma, \Lambda}_{z, x'}  $$
$$    \hat{O}_{x'}  =  C^{\gamma, \Lambda}_{y, x'}  \ \ \ \
    \hat{O}_{z'}  =  C^{\gamma, \Lambda}_{y, z'} ;  $$
and for TR:
$$  \hat{L}_{z'}  =  C^{p, \Lambda}_{z, z'} \ \ \ \
    \hat{L}_{x'}  =  C^{p, \Lambda}_{z, x'}  $$
$$    \hat{T}_{z'}  =  C^{p, \Lambda}_{x, z'}  \ \ \ \
    \hat{T}_{x'}  =  C^{p, \Lambda}_{x, x'} . $$ The Cartesian
components refer to the spin axis for the
baryons;  for the photon they indicate either
linear or circular polarization states (see Ref.~\cite{FTS}
for a discussion).

\section{Projection of Multipoles}

  Argand plots of the
the electric $E^{\pm}_\ell$
and magnetic $M^{\pm}_\ell$  multipoles
are obtained from the
CGLN amplitudes $F_1,F_2,F_3$ and $ F_4$
by the following projection integrals:
\begin{eqnarray}
 E^{+}_\ell &=& \frac{1}{2 (\ell+1)} \int^{1}_{-1} dx \big{\lbrace}
F_1 P_\ell - F_2 P_{\ell + 1}
\nonumber \\ &+& \frac{1}{\ell+1} (1 -x^2) F_3 P'_\ell +
\frac{1}{\ell+2} (1-x^2) F_4 P'_{\ell+1}    \big{\rbrace} \nonumber
\\
 E^{-}_\ell &=& \frac{1}{2 \ell} \int^{1}_{-1} dx \lbrace
-F_1 P_\ell + F_2 P_{\ell - 1} + \frac{1}{\ell} (1 -x^2) F_3 P'_\ell
  \nonumber \\ &+& \frac{1}{\ell-1} (1-x^2) F_4 P'_{\ell-1}
  \rbrace  \nonumber \\
 M^{+}_\ell &=& \frac{1}{2 (\ell+1)} \int^{1}_{-1} dx \lbrace
F_1 P_\ell - F_2 P_{\ell + 1}
\nonumber \\  &-& \frac{1}{\ell(\ell+1)} (1 -x^2) F_3 P'_\ell
 \rbrace  \nonumber
\\
 M^{-}_\ell &=& \frac{1}{2 \ell} \int^{1}_{-1} dx \lbrace
-F_1 P_\ell + F_2 P_{\ell - 1}
\nonumber \\ &+& \frac{1}{\ell(\ell+1)} (1 -x^2) F_3 P'_\ell
\rbrace \nonumber  .
\end{eqnarray}  Here $x=\cos \theta,$ $P_\ell$ are the Legendre
polynomials and the CGLN
amplitudes are given in Ref.~\cite{AS,WJC},
where they have been calculated using diagrammatic techniques in an
isobaric approach.

\end{appendix}

\newpage
  \begin{table}
\begin{tabular}{ccccccc}

AS&$E^+_0(1/2^-)$&$ \bbox {{ M}^-_1(1/2^+)}$&$ E^+_1 (3/2^+)$&$
E^-_2 (3/2^-)^\dagger$&$E^+_2(5/2^-)^\dagger$  \\
 & & &$   M^+_1(3/2^+)$&$M^-_2(3/2^-)^\dagger$&$M^+_2(5/2^-)^\dagger$ \\
WJC&$ \bbox {{ E}^+_0(1/2^-)}$&$E^+_1(3/2^+) $&
$\bbox {{ M}^-_1(1/2^+)}$&$ E^+_2(5/2^-)^\dagger$  &
\\
&&$  E^-_2(3/2^-)^\dagger$&$M^-_2(3/2^-)^\dagger $ &
$M^+_2(5/2^-)^\dagger$ &  \\
&&   &$M^+_1(3/2^+) $ & &  \\
SALY&$ E^+_0(1/2^-)$&$ \bbox {{M}^+_1(3/2^+)}$&$\bbox {{
M}^-_1(1/2^+)}$&
$\bbox {{ M}^-_2(3/2^-)}$&
$ M^+_2(5/2^-)^\dagger$ & \\
 & & &$ \bbox {{ E}^+_1(3/2^+)}$ & &$ E^+_2(5/2^-)^\dagger$ &  \\
 & & & $ \bbox {{ E}^-_2(3/2^-)}$   & & &
\end{tabular}
\caption{The multipoles for the three models listed in
order of their  maximum size from threshold to the maximum energy range
of the corresponding model,  based on Fig.~5.  The boldfaced multipoles
display counter-clockwise resonance looping.  The multipoles
marked by a dagger receive strength from {\it t-}channel exchange.}
\end{table}
\newpage
 \input epsf
\begin{figure}[b]
\caption{  The sixteen observables for kaon photoproduction
process $\gamma p \rightarrow K^+ \Lambda$, for the
AS(dashed), WJC(dotted) and SALY(solid) models are presented versus
kaon c.m. angle $\theta$ in degrees.
The photon laboratory energy is  $E_{\gamma}^{\rm lab} = 0.920 $ GeV,
which is just above the kaon production threshold of 0.911 GeV.
Observables of Legendre class
 ${\cal L}_0, {\cal L}_{1a}, {\cal L}_{1b}$ and ${\cal L}_{2}$
are presented in each column.  The first row gives the differential
cross section and the single spin
observables;
the next rows are the Beam-Target,
  the Beam-Recoil and the Target-Recoil double spin observables.
}
\end{figure}

\begin{figure}[b]
\centerline{
\epsfxsize=4.0in
 }
\caption{  Same as Fig.~1, but for a photon energy of
 $E_{\gamma}^{\rm lab} = 1.4$ GeV.
}
\end{figure}

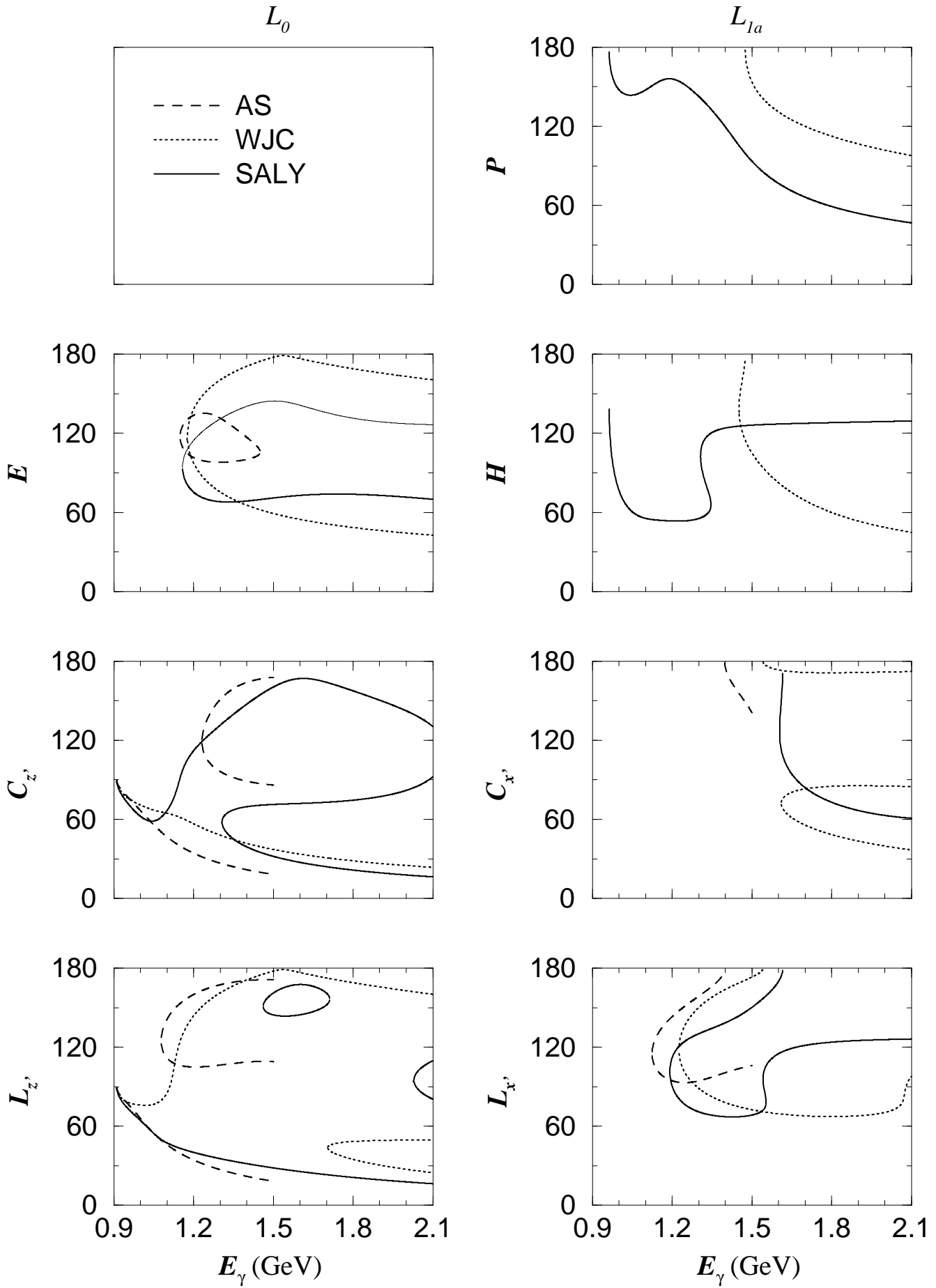
\begin{figure}[b]
\centerline{
\epsfxsize=4.0in
 }
\caption{  The nodal trajectory plots for the
spin observables of Legendre class:
 (a) ${\cal L}_0,$ (b) ${\cal L}_{1a},$ (c) ${\cal L}_{1b}$
 and (d) ${\cal L}_2.$  The c.m. kaon angle at which a sign-changing
zero occurs for a given spin observable is plotted versus
the incident laboratory photon energy. The single spin observables are
presented in the top row, while
the Beam-Target, Beam-Recoil and Target-Recoil double
spin observables are located in the
second, third and bottom row, respectively. Again the curves are
displayed as AS(dashed), WJC(dotted) and SALY(solid).
}
\end{figure}
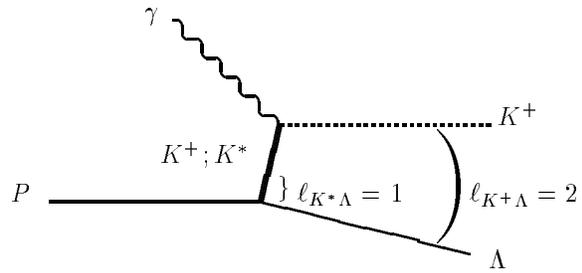
\begin{figure}[b]
 \centerline{
 \epsfxsize=4.0in
 }
\caption{ The {\it t-}channel or kaon exchange mechanism.  The virtual
kaon $(K^+ {\rm or\ } K^*   )$ is typically emitted in a $P-$wave.
 The incident photon can boost the virtual kaon
to a $\ell_{K^+\Lambda} =2$ state.  Then the final state  angular
momentum is obtained by adding in the $\Lambda$ spin
as $\vec{2} + \vec{1/2}\rightarrow 3/2^-, 5/2^-;$  the negative parity
arises from the rule $-(-1)^\ell.$  This mechanism feeds into the
$(E^-_2 M^-_2)3/2^-$ and $(E^+_2 M^+_2)5/2^-$ multipoles.  It is also
possible for the incident photon to lower $\ell_{K^+\Lambda}$ to zero,
  which could
affect the $E^+_0$ $1/2^-$ amplitudes;
  this multipole is already large so the major effect is the one
illustrated here.  If the kaon is left in a $P-$wave after the photon
is absorbed,  then the above {\it t-}channel
can contribute to the $3/2^+$   $ M_1^+$ and $1/2^+$ $ E_1^-, M_1^- $
states;  however,  these multipoles are usually dominated by
{\it s-}channel contributions.  If the virtual kaon is the $K1,$ it
is produced
mainly in an $S-$wave and therefore yields $\ell_{K^+ \Lambda}=1,0$ or
$J= 1/2^+ , 3/2^+$ or $1/2^-$ states,  which are usually dominated by
{\it s-}channel contributions.
}
\end{figure}
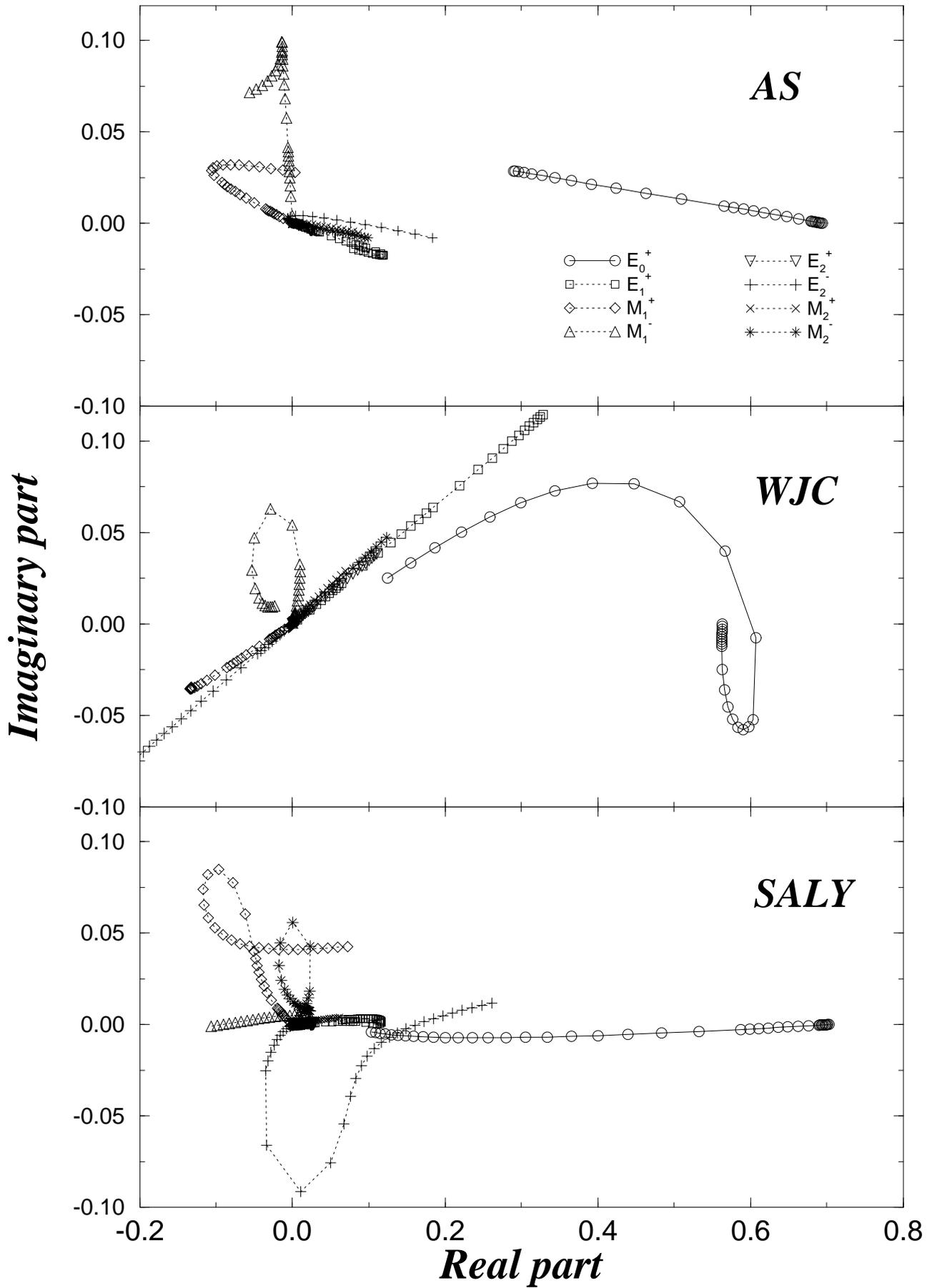
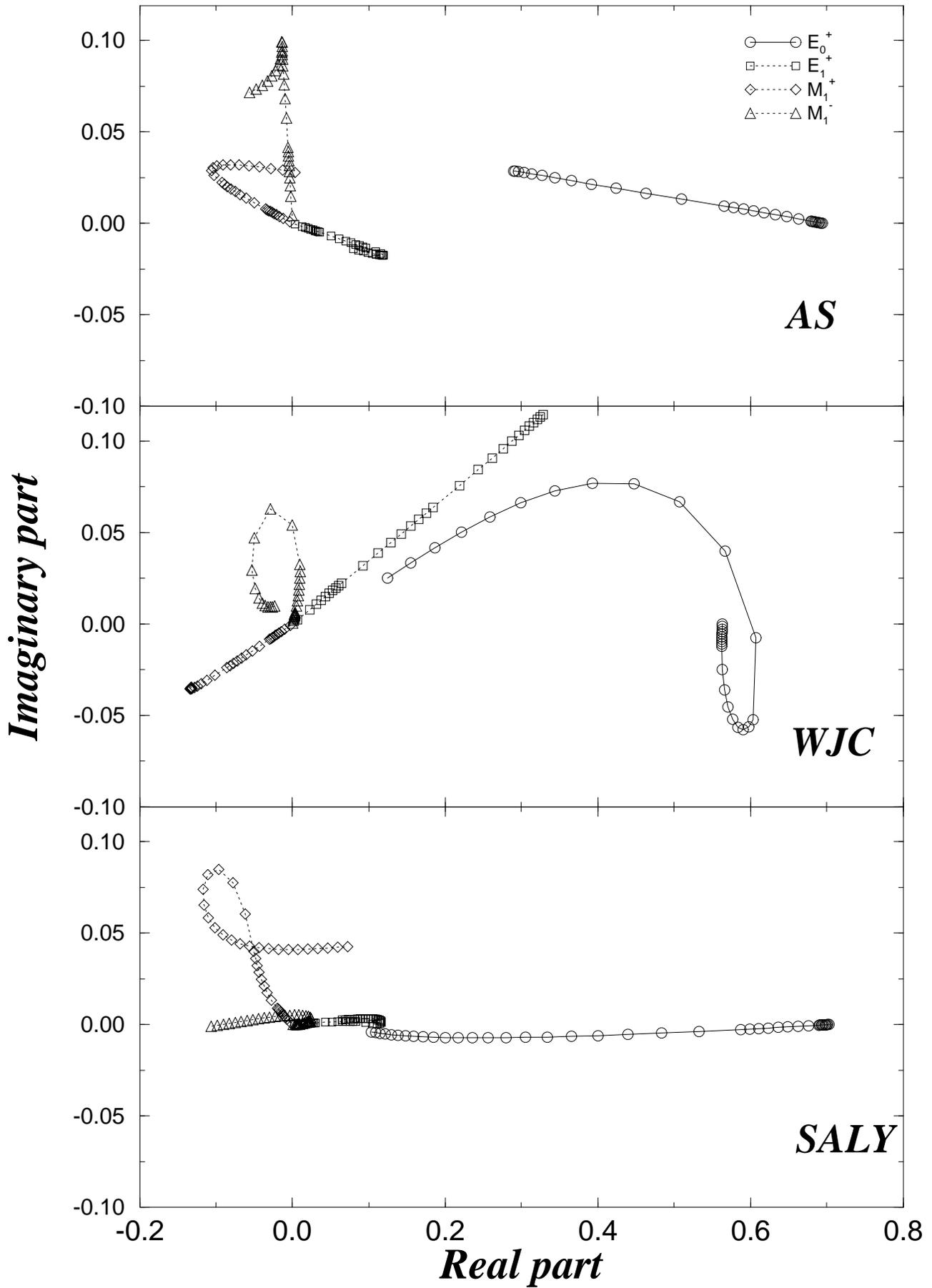
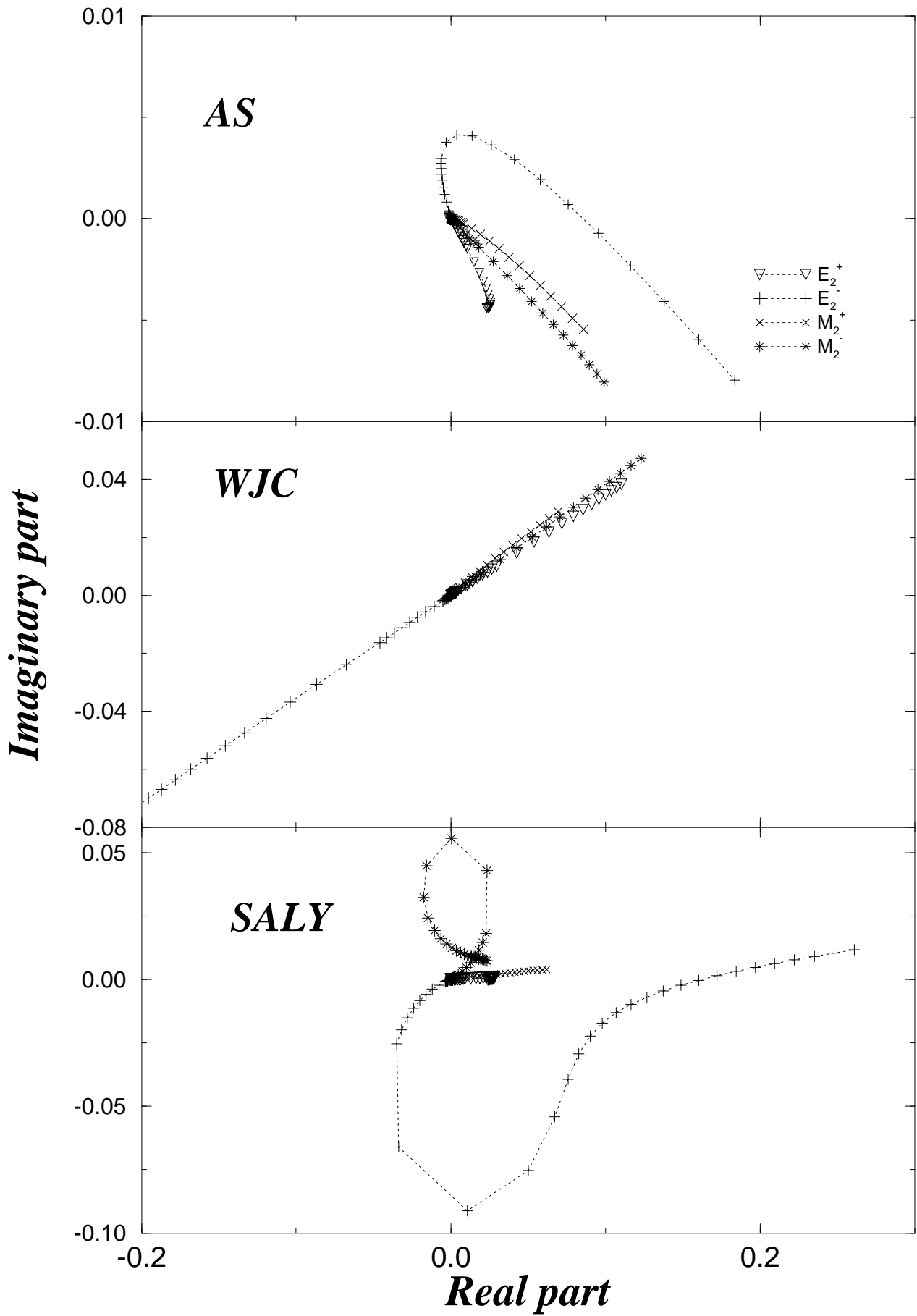
\begin{figure}[b]
 \centerline{
 \epsfxsize=4.0in
 }
\caption{ Argand plots of the electric and magnetic multipoles for the
 AS, WJC and SALY models for: (a) $\ell \leq 1$ and (b) $\ell =2.$
   The S-wave multipole $E^+_0$
starts at a nonzero point along the real axis and then evolves with
energy.  All other multipoles start at the origin and follow the
trajectories shown as energy increases. For resonant states the Argand
plot exhibits the usual looping.
The scale in (a) is kept fixed,  but it varies with model in (b).
Note the role of these amplitudes in observables is typically
 weighted by
$2J+1$ factors;  hence $D-$wave effects can be magnified.
}
\end{figure}

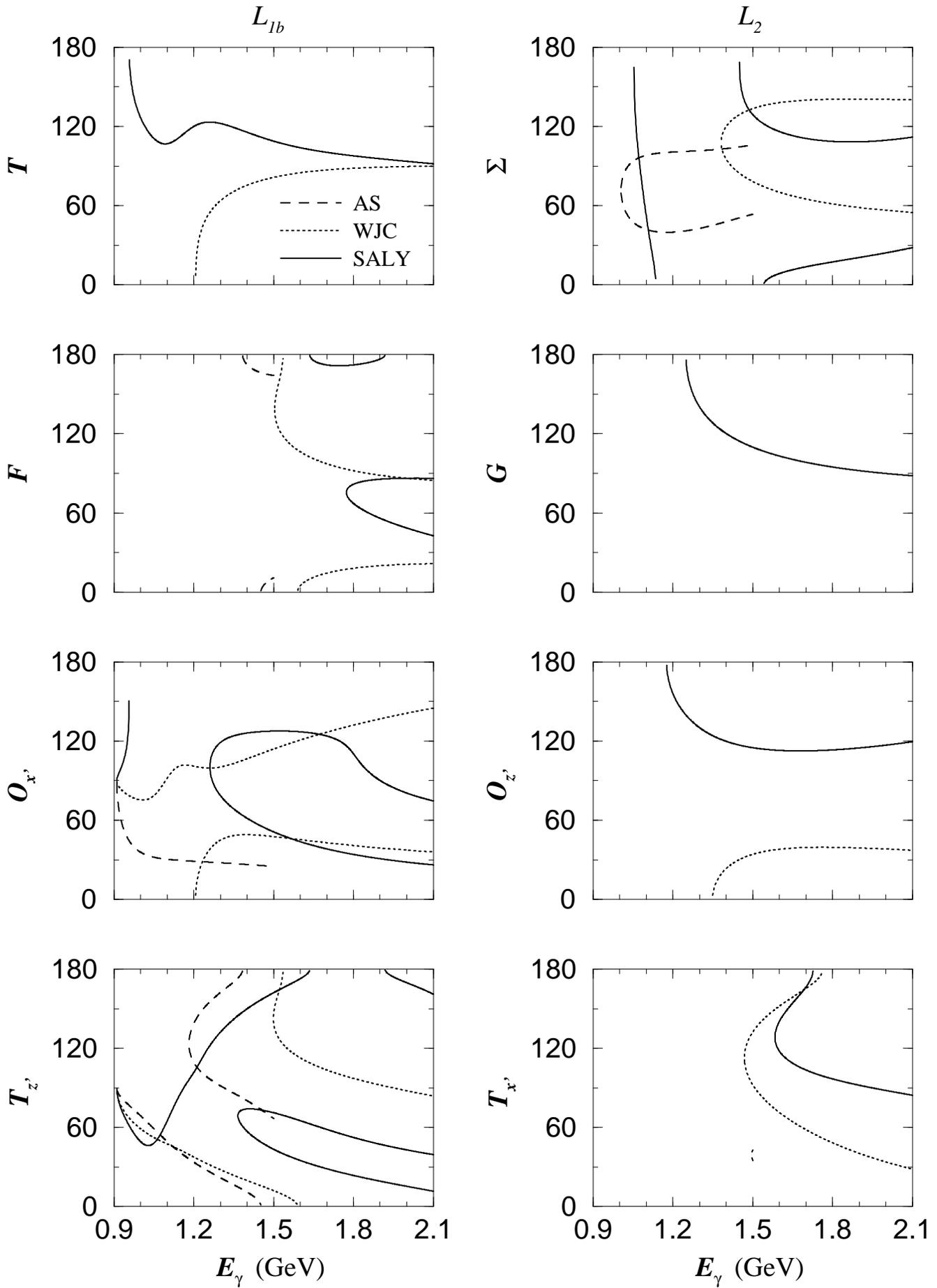
\end{document}